\documentclass{article}
\usepackage{emulateapj,epsfig}

\slugcomment{Revised, accepted for publication in ApJ Letters, May 2, 2000}

\makeatletter
\newenvironment{inlinetable}{%
\def\@captype{table}%
\noindent\begin{minipage}{0.999\linewidth}\begin{center}\footnotesize}
{\end{center}\end{minipage}\smallskip}

\newenvironment{inlinefigure}{%
\def\@captype{figure}%
\noindent\begin{minipage}{0.999\linewidth}\begin{center}}
{\end{center}\end{minipage}\smallskip}
\makeatother

\begin{document}


\title{COLLISIONAL DARK MATTER AND THE STRUCTURE OF DARK HALOS}


\author{Naoki Yoshida\altaffilmark{1}, Volker
Springel\altaffilmark{1,2}, Simon D.M. White\altaffilmark{1}}
\affil{Max-Planck-Institut f\"{u}r Astrophysik Karl-Schwarzschild-Str. 1, 85748 Garching, Germany}
\and
\author{Giuseppe Tormen\altaffilmark{3}}
\affil{Dipartimento di Astronomia, Universita di Padova,
vicolo dell'Osservatorio 5, 1-35122 Padova, Italy}

\altaffiltext{2}{present address: Harvard-Smithsonian Center for Astrophysics,
    60 Garden Street, Cambridge, MA 02138}

\begin{abstract}
We study how the internal structure of dark halos is affected if Cold
Dark Matter particles are assumed to have a large cross-section for 
elastic collisions. We identify a cluster halo in a large cosmological 
N-body simulation and resimulate its formation with progressively 
increasing resolution. We compare the structure found in the two cases
where dark matter is treated as collisionless or as a fluid. For the
collisionless case the overall ellipticity of the cluster, the central
density cusp and the amount of surviving substructure are all similar
to those found in earlier high resolution simulations. Collisional
dark matter results in a cluster which is more nearly spherical at all
radii, has a {\it steeper} central density cusp, and has less, but still
substantial surviving substructure. As in the colisionless case, these
results for a ``fluid'' cluster halo are expected to carry over
approximately to smaller mass systems. The observed rotation curves of
dwarf galaxies then argue that self-interacting dark matter can only 
be viable if intermediate cross-sections produce structure which does
not lie between the extremes we have simulated.
\end{abstract}




\section{Introduction}
Cold dark matter scenarios within the standard inflationary
universe have proved remarkably successful in fitting a wide range of
observations. While structure on large scales is well 
reproduced by the models, the situation is more controversial in the
highly nonlinear regime. Navarro, Frenk \& White (1995, 1996, 1997;
NFW) claimed that the density profiles of near-equilibrium dark halos 
can be approximated by a ``universal'' form with singular 
behaviour at small radii. Higher resolution studies have confirmed 
this result, finding even more concentrated dark halos than the original
NFW work and showing, in addition, that CDM halos are predicted to have
a very rich substructure with of order 10\% of their mass contained
in a host of small subhalos (Frenk et al 1999, Moore et al 1999a, 1999b,
Ghigna et al 1999, Klypin et al 1999, Gottloeber et al 1999, White \& 
Springel 1999). Except for a weak anticorrelation of concentration with 
mass, small and large mass halos are found to have similar structure.
Many of these studies note that the predicted concentrations appear 
inconsistent with published data on the rotation curves of dwarf
galaxies, and that the amount of substructure exceeds that seen in 
the halo of the Milky Way (see also Moore 1994; Flores and Primack 1994;
Kravtsov et al 1998; Navarro 1998).

It is unclear whether these discrepancies reflect a fundamental
problem with the Cold Dark Matter picture, or are caused
by overly naive interpretation of the observations of the
galaxy formation process (see Eke, Navarro \&
Frenk 1998; Navarro \& Steinmetz 1999; van den Bosch 1999). On the
assumption that an explanation should be sought in fundamental
physics, Spergel \& Steinhardt (1999) have argued that a large
cross-section for elastic collisions between CDM particles may
reconcile data and theory. They suggest a number of modifications
of standard particle physics models which could give rise to such 
self-interacting dark matter, and claim that cross-sections which
lead to a transition between collisional and collisionless
behaviour at radii of order 10 -- 100 kpc in galaxy halos are
preferred on astrophysical grounds. Ostriker (1999) argues that
the massive black holes observed at the centres of many galactic
spheroids may arise from the accretion of such collisional dark matter
onto stellar mass seeds. Miralda-Escude (2000) argues that such dark
matter
will produce galaxy clusters which are rounder than observed and so can
be
excluded.

At early times the CDM distribution is indeed cold, so
the evolution of structure is independent of the collision
cross-section of the CDM particles. At late times, however, a large
cross-section leads to a small mean free path and so to fluid 
behaviour in collapsed regions. In this Letter we explore
how the structure of nonlinear objects (``dark halos'') is affected
by this change. We simulate the formation of a massive halo from
CDM initial conditions in two limits: purely collisionless dark
matter and ``fluid'' dark matter. We do not try to simulate the
the more complex intermediate case in which the mean free path
is large in the outer regions of halos but small in their cores.
If this intermediate case (which is the one favoured by Spergel 
\& Steinhardt (1999) and by Ostriker (1999)) produces nonlinear 
structure intermediate between the two extremes we do treat, then 
our results show that collisional CDM would give poorer fits to 
the rotation curves of dwarf galaxies than standard collisionless 
CDM. Further work is needed to see if this is indeed the case.

\section{THE N-BODY/SPH SIMULATION}
Our simulations use the parallel tree code GADGET developed by
Springel (1999, see also Springel, Yoshida \& White 2000b). 
Our chosen halo is the second most massive cluster
in the $\Lambda$CDM simulation of Kauffmann et al (1999). We analyse
its structure in the original simulation and in two higher resolution
resimulations. In the collisionless case these are the lowest 
resolution members of a set of four resimulations carried out by
Springel et al (2000a) using similar techniques to those of NFW.
Details may be found there and in Springel et al(2000b). 
These collisionless resimulations use GADGET as an N-body solver, 
whereas our collisional
resimulations start from identical initial conditions but use the code's
Smoothed Particle Hydrodynamics (SPH) capability to solve the fluid
equations. The SPH method regards each simulation particle as a
``cloud'' of fluid with a certain kernel shape. These clouds interact
with each other over a length scale which is determined by the local 
density and so varies both in space and time.
The basic parameters of our simulations are tabulated in Table 1,
where N$_{\mbox{\small{tot}}}$ is the total number of particles in the
simulation,
N$_{\mbox{\small{high}}}$ the number of particles in the central
high-resolution
region, $m_{p}$ is the mass of each high-resolution particle, and
$l_{s}$ stands for the gravitational softening length.
Our cosmological model is flat with matter density $\Omega_{m}=0.3$,
cosmological constant $\Omega_{\Lambda}=0.7$ and expansion rate
$H_{0}=70$km$^{-1}$Mpc$^{-1}$.  It has a CDM power spectrum normalised
so that $\sigma_{8}=0.9$. The virial mass of the final cluster is
$M_{200}=7.4\times 10^{14}h^{-1}M_\odot$, determined as the mass within
the radius $R_{200}= 1.46 h^{-1}$Mpc where the enclosed mean

\begin{inlinetable}
\vspace*{0.4cm}\ \\
\begin{center}
\caption{Simulation parameters}
\begin{tabular}{ccccc}
\tableline
\tableline
Run & $N_{\rm tot}$ & $N_{\rm high}$ & $m_{\rm p}$ ($h^{-1} M_{\odot}$)
& $l_{\rm s}$($h^{-1}$kpc) \\
\tableline
S0 & 3.2$\times 10^{6}$ & 0.2$\times 10^{6}$& 1.4 $\times 10^{10}$ & 30 \\
\tableline
S1 & 3.5$\times 10^{6}$ & 0.5$\times 10^{6}$& 0.68 $\times 10^{10}$ & 20\\
\tableline
S2 & 5.1$\times 10^{6}$ & 2.0$\times 10^{6}$& 0.14 $\times 10^{10}$ & 3.0\\
\tableline
\end{tabular}
\end{center}
\end{inlinetable}

\section{RESULTS}
On scales larger than the final cluster, the matter distribution
in all our simulations looks similar. This is no surprise. The initial
conditions in each pair of simulations are identical, so
particle motions only begin to differ once pressure forces become
important. Furthermore the initial perturbation fields in simulations 
of differing resolution are identical on all scales resolved in both
models, and even S0 resolves structure down to scales well below that
of the cluster. As is seen clearly in Figure 1, a major difference 
between the collisional and collisionless models is that the final
cluster
is nearly spherical in the former case and quite elongated in the
latter. The axial ratios determined from the inertia tensors of the 
matter at densities exceeding 100 times the critical value are 
1.00:0.96:0.84 and 1.00:0.72:0.63 respectively. Again this
is no surprise. A slowly rotating fluid body in hydrostatic 
equilibrium is required to be nearly spherical, but no such constraint
applies in the collisionless case (see also Miralda-Escude 2000).

\begin{inlinefigure}
\vspace*{0.2cm}\ \\
\resizebox{8.5cm}{!}{\includegraphics{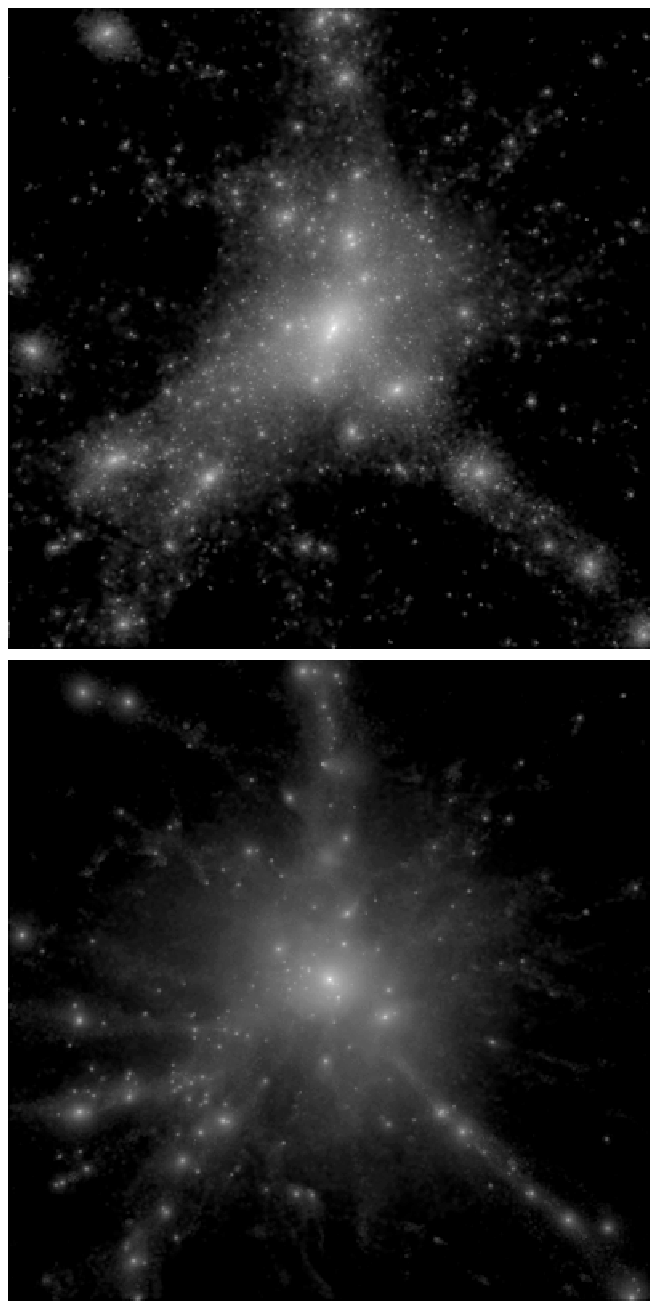}}
\caption{The projected mass distribution in our two
highest resolution simulations. The collisionless case (S2) is on
the top and the fluid case (S2F) is on the bottom. The region shown
is a cube of $15h^{-1}$Mpc on a side. \label{fig1}}
\end{inlinefigure}

In Figure 2 we show circular velocity profiles for our simulations.
These are defined as $V_c(r) = \sqrt{GM(r)/r}$, where $M(r)$ is the mass 
within a sphere radius $r$; they are plotted at radii between 2$l_{s}$
and $5R_{200}$. 
They agree reasonably well along each sequence of increasing
resolution,
showing that our results have converged numerically on these scales.
Along the fluid sequence the profiles resemble the collisionless
case over the bulk of the cluster. In the core, 
however, there is a substantial and significant difference; the fluid 
cluster has a substantially steeper central cusp. The difference
extends out to radii of about $0.5R_{200}$ and has the wrong
sign to improve the fit of CDM halos to published rotation curves for
dwarf and low surface brightness galaxies.

\begin{inlinefigure}
\vspace*{0.2cm}\ \\
\resizebox{8cm}{!}{\includegraphics{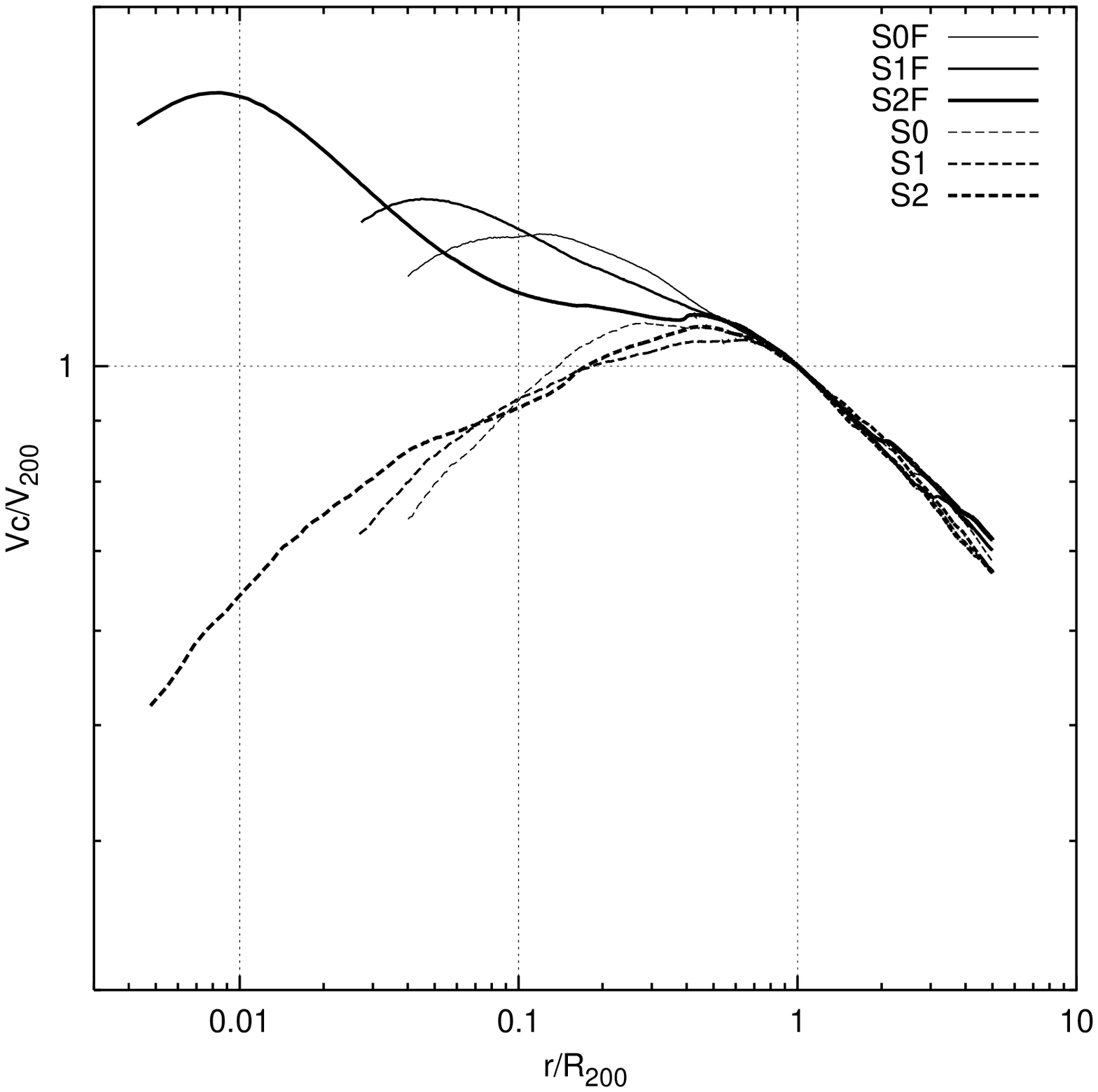}}
\caption[fig2.ps]{Circular velocity profiles for our cluster
simulations, each normalized to its own $R_{200}$ and $V_{200}$. These are
plotted between twice the gravitational softening and $5R_{200}$. The collisionless sequence is
plotted using dashed lines and the fluid sequence using solid lines.
\label{fig2}}
\end{inlinefigure}

(Note that in the fluid
case we expect small halos to approximate scaled down but slightly
{\it more} concentrated versions of cluster halos, as in the
collisionless case studied by Moore et al (1999a); this scaling will
{\it fail} for intermediate cross-sections because the ratio of the
typical mean free path to the size of the halo will increase with halo
mass.)

In Figure 3 we compare the level of substructure within $R_{200}$
in our various simulations. Subhalos are identified using
the algorithm SUBFIND by Springel (1999) which defines them
as maximal, simply connected, gravitationally self-bound sets 
of particles which are at higher local density than all surrounding 
cluster material. (Our SPH scheme defines a local density in the 
neighbourhood of every particle.) 
Using this procedure we find
that 1.0\%, 3.4\% and 6.7\% of the mass within $R_{200}$ is included
in subhalos in S0, S1 and S2 respectively. Along the fluid sequence
the corresponding numbers are 3.0\%, 6.4\% and 3.1\%.  
The difference in the total amount results primarily from the 
chance inclusion or exclusion of infalling massive
halos near the boundary at $R_{200}$.
In Figure 3 we show the mass distributions of
these subhalos. We plot each simulation to a mass limit of 40
particles, corresponding approximately to the smallest structures we 
expect to be adequately resolved in our SPH simulations. Along each
resolution sequence the agreement is quite good, showing this limit to
be conservative. For small subhalo masses there is clearly less 
substructure in the fluid case, but the difference is
more modest than might have been anticipated.

\section{Summary and Discussion}
An interesting question arising from our results is {\it why} our
fluid clusters have more concentrated cores than their collisionless 
counterparts. The density profile of an equilibrium gas sphere can be
thought of as being determined by its Lagrangian specific entropy
profile, i.e. by the function $m(s)$ defined to be the mass of gas
with specific entropy less than $s$. The larger the mass at low
specific entropy, the more concentrated the resulting profile. Thus
our fluid clusters have more low entropy gas than if their profiles
were similar to those of the collisionless clusters. The entropy of the
gas is produced by a variety of accretion and
merger shocks during the build-up of the cluster, so the strong central
concentration reflects a relatively large amount of weakly
shocked gas.

\begin{inlinefigure}
\vspace*{0.3cm}\ \\
\resizebox{8cm}{!}{\includegraphics{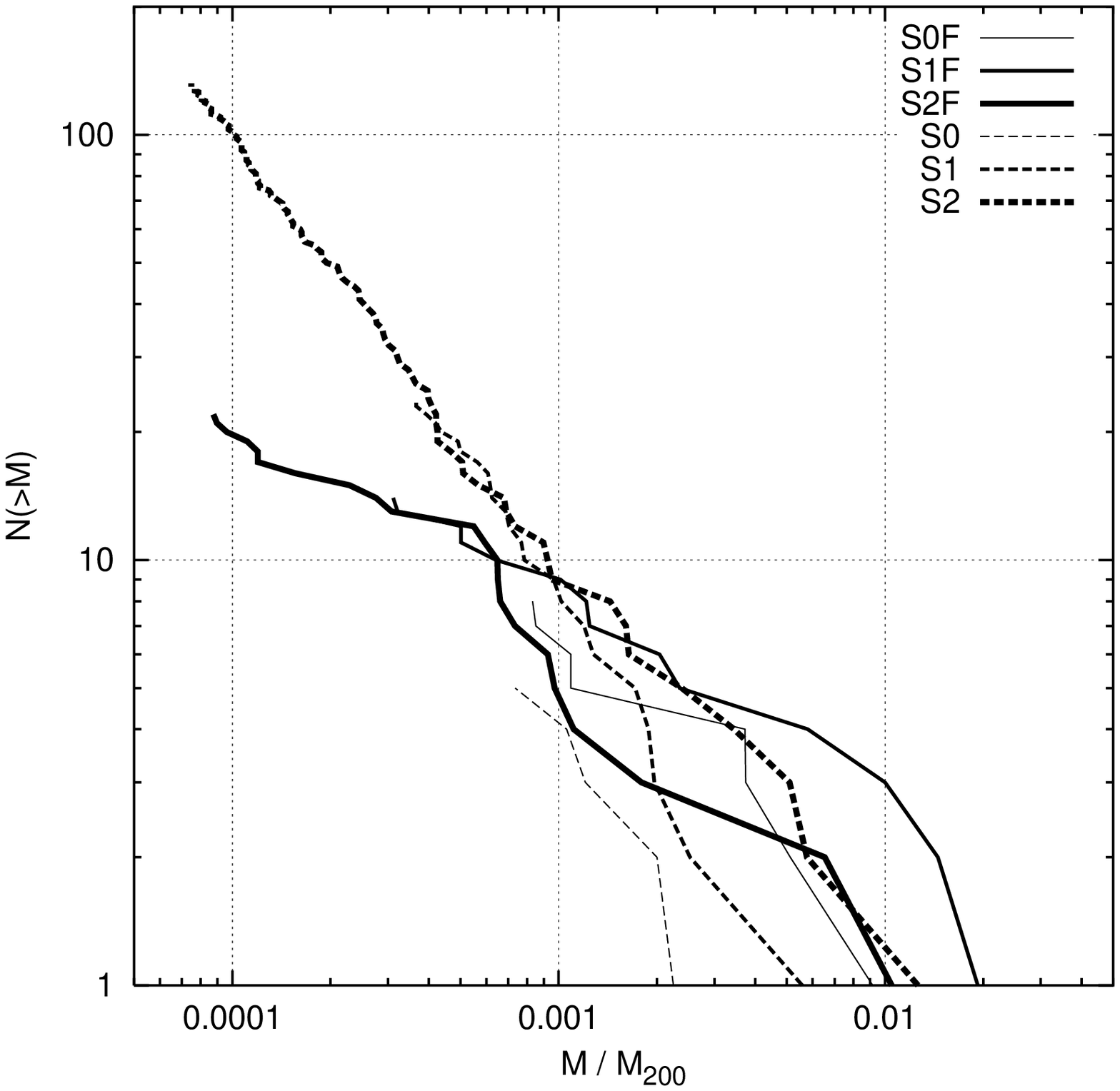}}
\caption{The total number of subhalos within $R_{200}$ is
plotted as a function of their mass in units of $M_{200}$.
Dashed and solid lines correspond to the collisionless 
and fluid cases respectively. Results for each simulation are 
plotted only for halos containing more than 40 particles.
\label{fig3}}
\end{inlinefigure}

We study gas shocking in our models by carrying out
one further simulation. We take the initial conditions of S1 and
replace each particle by two superposed particles, a collisionless
dark matter particle containing 95\% of the original mass and a gas
particle
containing 5\%. These two then move together until SPH pressure forces
are strong enough to separate them. The situation is similar to the
standard 2-component model for galaxy clusters except that our chosen
gas fraction is significantly smaller than observed values.

In this mixed simulation the evolution of the collisionless matter
(and its final density profile) is almost identical to that in the
original S1. This is, of course, a consequence of the small gas
fraction we have assumed. In agreement with the simulations
in Frenk et al (1999) we find that the gas density profile parallels
that of the dark matter over most of the cluster but is significantly
{\it shallower} in the inner $\sim 200 h^{-1}$kpc.  Comparing this new
simulation (S1M) with its fluid counterpart (S1F) we find that in both
cases the gas which ends up near the cluster centre lay
at the centre of the most massive cluster progenitors at $z= 1\sim 3$.
In addition it is distributed in a similar way among the progenitors 
in the two cases. In Figure 4 we compare the specific entropy profiles
of the cluster gas. These are scaled so that they would be identical 
if each gas particle had the same shock history in the two simulations. 
Over most of the cluster there is indeed a close correspondence,
but near the centre the gas in the mixed simulation has higher
entropy. (This corresponds roughly to $r < 100h^{-1}$kpc.)

\begin{inlinefigure}
\vspace*{0.3cm}\ \\
\resizebox{8cm}{!}{\includegraphics{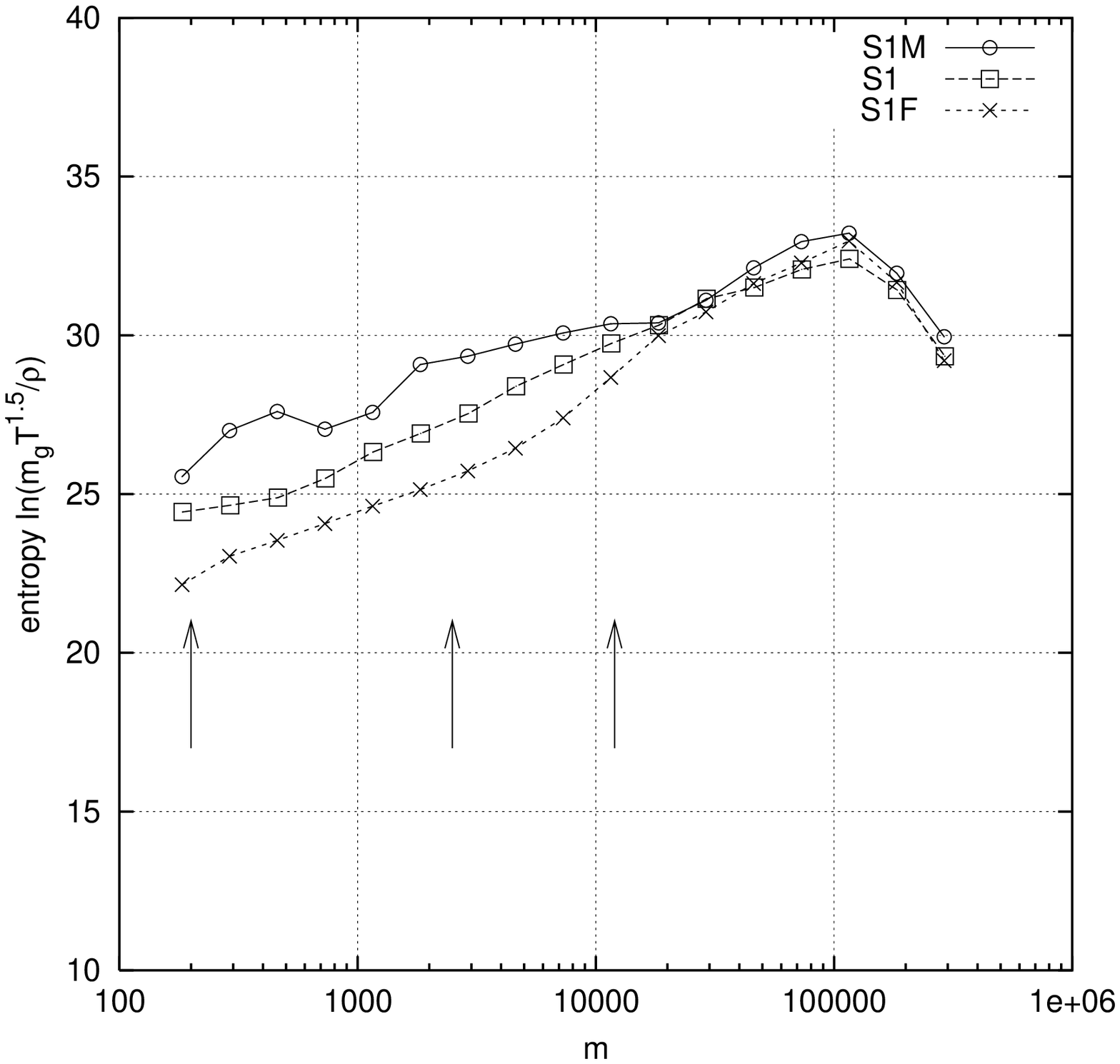}}
\caption{We plot Lagrangian specific entropy profiles for
the gas fluid simulation (S1F: crosses) and for the mixed simulation
(S1M: open circles). In each case $m(s)$ is given in units 
of the individual gas particle mass, $m_g$, and the specific entropy
of a particle is defined as $\ln (m_g T_g^{1.5}/\rho_g)$. The 
arrows indicate where the
timescale $t_{2b}$ for 2-body heating of the gas by encounters with dark matter
particles (see equation (5) of Steinmetz \& White (1997)) is 0.1, 1,
and 10 times the age 
of the Universe. For each $s$ we calculate $t_{2b}$ 
at the radius where the median specific 
entropy equals $s$. The dashed line with open squares is an
``entropy'' profile for S1 calculated by using the
SPH kernel to calculate the density and velocity dispersion in the
neighborhood of each particle, and then converting from velocity dispersion
to temperature using the standard relation for a perfect monatomic
gas.
\label{fig4}}
\end{inlinefigure}

As Figure 4 shows, this is partly a numerical artifact; the two 
entropies differ only at radii where two-body heating of the gas by
the dark matter particles is predicted to be important in the mixed
case. (The effect is absent in the pure fluid simulation.)
The weaker shocking in the fluid case is
evident from the equivalent "entropy" profile of S1 in Figure 4. This
lies between those of the two fluid simulations, and in particular
significantly above that of S1F in the central regions.

In conclusion the effective heating of gas by shocks in the fluid case
is similar to but slightly weaker than that in the mixed case. This is
presumably a reflection of the fact that the detailed morphology
of the evolution also corresponds closely. The difference in 
final density profile is a consequence of three effects. In the mixed 
case the gas is in equilibrium within the external potential 
generated by the dark matter, whereas in the pure fluid case it 
must find a self-consistent equilibrium. In addition the
core gas is heated by two-body effects in the mixed case. Finally
in the pure fluid case the core gas experiences weaker shocks.

Overall our results show that in the large cross-section limit 
collisional dark matter is not a promising candidate for improving 
the agreement between the predicted structure of CDM halos and 
published data on galaxies and galaxy clusters. The increased 
concentration at halo centre will worsen the apparent conflict 
with dwarf galaxy rotation curves. Furthermore, clusters are predicted
to be nearly spherical and galaxy halos to have similar mass in 
substructure to the collisionless case, although with fewer low 
mass subhalos. Intermediate
cross-sections would lead to collisional behaviour in dense regions 
and collisionless behaviour in low density regions with a consequent
breaking of the approximate scaling between high and low mass halos.
The resulting structure may not lie between the two extremes
we have simulated. Self-interacting dark matter might then help
resolve the problems with halo structure in CDM models, if indeed 
these problems turn out to be real rather than apparent.

\acknowledgments
SW thanks Jerry Ostriker and
Mike Turner for stimulating discussions which started him thinking
about this project.






\end{document}